\documentclass[superscriptaddress,secnumarabic,
amssymb,amsmath,nobibnotes,aps,prd,showkeys,showpacs,nofootinbib]{revtex4}%
\usepackage{graphicx}
\usepackage{epsf}
\usepackage{bm}
\usepackage{amsmath}
\usepackage{amsfonts}
\usepackage{amssymb}
\usepackage{epstopdf}
\usepackage{subfigure}
\usepackage{natbib}
\usepackage{color}%
\usepackage{hyperref}
\hypersetup{dvips,dvipdfm,linktoc=page,colorlinks=true,linkcolor=blue,citecolor=red,filecolor=magenta,urlcolor=magenta,bookmarks=true}
\providecommand{\U}[1]{\protect\rule{.1in}{.1in}}
\newcommand{\be}{\begin{equation}}
\newcommand{\ee}{\end{equation}}

\newcommand{\mincir}{\raise
-3.truept\hbox{\rlap{\hbox{$\sim$}}\raise4.truept\hbox{$<$}\ }}
\newcommand{\magcir}{\raise
-3.truept\hbox{\rlap{\hbox{$\sim$}}\raise4.truept\hbox{$>$}\ }}

\begin{document}
\title{ Dynamical analysis of interacting dark energy model in the framework of particle creation mechanism }
\author{Sujay Kr. Biswas}
\email{sujaymathju@gmail.com}
\affiliation{Department of Mathematics, Ramakrishna Mission Vivekananda Centenary College, Rahara, Kolkata-700 118, West Bengal, India.}
\affiliation{Department of Mathematics, Jadavpur University, Jadavpur, Kolkata - 700032, West Bengal, India}
\author{Wompherdeiki Khyllep}
\email{sjwomkhyllep@gmail.com}
\affiliation{ Department of Mathematics, St. Anthony's College, Shillong- 793001, Meghalaya, India}
\author{Jibitesh Dutta}
\email{jdutta29@gmail.com,jibitesh@nehu.ac.in}
\affiliation{ Mathematics Division, Department of Basic
Sciences and Social Sciences,~ North Eastern Hill University,~NEHU
Campus, Shillong - 793022, Meghalaya, India}
\author{Subenoy Chakraborty}
\email{schakraborty@math.jdvu.ac.in} \affiliation{Department of
Mathematics, Jadavpur University, Jadavpur, Kolkata - 700032, West
Bengal, India} 
\begin{abstract}
In this work we present the cosmological dynamics of interacting
dark energy models in the framework of particle creation
mechanism. The particle creation mechanism presented here
describes the true non equilibrium thermodynamics of the universe.
In spatially flat Friedmann-Lema{\^\i}tre-Robertson-Walker
universe considered here, the dissipative bulk viscous pressure is
due to the non conservation of particle number. For simplicity, we
assume that the creation of perfect fluid particles to be
isentropic (adiabatic) and consequently the viscous pressure obeys
a linear relationship with particle creation rate. Due to
complicated nature of the Einstein's field equations, dynamical
systems analysis has been performed to understand deeply about the
cosmological dynamics. We have found some interesting cosmological
scenarios like late-time evolution of universe dominated by dark
energy which could mimic quintessence, cosmological constant or
phantom field through a dark matter dominated era. We also
obtained a possibility of crossing the phantom divide line which
is favored by observations.
\end{abstract}
\maketitle
\section{Introduction}

 Various observations  suggest that our universe is currently
undergoing a phase of accelerated expansion
\cite{Riess:1998cb,Perlmutter:1998np,Betoule:2014frx,Ade:2013zuv,Ade:2015xua}.
This is a challenging issue in standard cosmology which shows a
new imbalance in the governing Friedmann equations. People have
addressed such imbalances either by introducing new sources or by
altering the governing equations. In the frame of standard
cosmology, the first one is termed as dark energy with a huge
negative pressure, and the second one involves the introduction of
some modifications into the gravity sector commonly known as
modified gravity theories.  The simplest dark energy (DE)
candidate is the cosmological constant $\Lambda$, which together
with cold dark matter provides the simplest cosmological model
known as the $\Lambda$-cold-dark-matter ($\Lambda$CDM), which
according to a large number of observations, is the best
cosmological model at present. However, $\Lambda$CDM suffers from
severe problems in the interface of cosmology and particle
physics, such as  the cosmological constant problem \cite{S.
Weinberg1989, V. Sahni2000, T. Padmanabhan2003}  and the cosmic
coincidence problem \cite{I. Zlatev1999}.

In order to address these issues related to $\Lambda$-cosmology,
an extensive analysis have been performed ranging from
 various DE models to modified gravity theories \cite{E. Copeland2006, L. Amendola2010}. Amongst them, the cosmological models where dark matter (DM) and
 DE interacts with each other have gained significant attention with the successive number of observational data. Although latest observations indicate a nonvanishing interaction
 in the dark sector \cite{int1, nps,kn}, but this interaction is very compatible to zero within the $1\sigma$ confidence region. In any case, the interaction between DE and DM could be a major issue to be confronted in studying the physics of DE.
However, since the nature of these two dark components (DE and DM)
remaining unknown, so the precise form of the  interaction is
unknown till now, and as such, there is no fundamental theory for
choosing a specific coupling. So, the choice of coupling is purely
phenomenological.
 Further, in the framework of field theory, it is natural to consider the inevitable interaction between the dark
 components. Interacting dark sector models have been extensively studied in
 several works \cite{N.Tamanini2015, C.G.Bohmer2008, Xi-ming Chen2009, T.Harko2013, Yuri.L.Bolotin2014, Andre A.Costa2014, M.Khurshudyan2015, S.Kr.Biswas2015a, S.Kr.Biswas2015b, L.P.Chimento2010, L.P.Chimento2012, J.S.Wang2015, S.Pan2015mnras}.
 In fact, an appropriate interaction between DE and DM can provide a mechanism to alleviate the coincidence problem \cite{L.P.Chimento2003} and cosmic age problem. Furthermore,
  it also provides a possibility of crossing  the phantom divide line \cite{S.Das2006, S.Pan2014} and explains the transient nature of the deceleration parameter.
  It should be noted that there are other options apart from the above mentioned choices
   for explaining the cosmic coincidence and other cosmological conundrums. In particular, there is the $\Lambda$XCDM type of models (for detailed study see
    \cite{Sola1,Sola2}) where there exists an interaction between the vacuum energy and another DE component (X). In this case, matter can be conserved and nevertheless the ratio
     between the DE and DM
  remains bounded in the entire cosmic history. Further, in this context one finds effective quintessence and phantom like behaviors in references \cite{Sola3,Sola4,Sola5}.

Therefore, interacting DE models provide richer cosmological
dynamics than non-interacting one by allowing the energy exchange
between dark sectors. This might provides a similar energy density
in dark sectors which can be achieved by accelerated scaling
attractor solution \cite{C.Wetterich1995, L.Amendola1999,
C.G.Bohmer2008} with

\begin{equation}\label{coincidence}
 \frac{\Omega_{\rm DE}}{\Omega_{\rm DM}}\approx \mathcal{O}(1)~~\mbox{and}~~ \omega_{\rm eff}<-\frac{1}{3}
\end{equation}
Thus, the proper choice of parameters without fine-tuning of the
initial conditions is required in order to match the ratio of
energy densities of dark sectors with observations.

Present observations \cite{Ade:2013zuv, Ade:2015xua, A. Rest,
J.-Q. Xia, C. Cheng, D.L. Shafer} also favors the possibility that
our universe entering into the phantom era with an effective
equation of state $\omega_{\rm eff}<-1$. To obtain this scenario,
usually scalar field with negative kinetic term is introduced
\cite{R.R. Caldwell2002}. However, this leads to some
instabilities at classical as well as quantum levels \cite{S.M.
Carroll2003, J.M. Cline2004} and also induces some other
theoretical problems \cite{S.D.H. Hsu2004, F. Sbisa2015, M.
Dabrowski2015}.

Another choice to explain this present acceleration is the
particle creation mechanism. This model can successfully mimic the
$\Lambda$CDM cosmology \cite{G.Steigman2009, J.A.S.Lima2010,
J.A.S.Lima2014, J.C.Fabris2014, S.Chakraborty2015}. Historically,
in 1939, Schrodinger \cite{E.Schrodinger1939} introduced a
microscopic description of particle production in an expanding
universe where gravity plays a crucial role. Following his idea,
Parker with collaborators \cite{L. Parker and collaborates} and
Zeldovich with his collaborators \cite{Zeldovich and others}
started investigating the possible physical scenarios by the
production of particles. Since the evolution of universe could be
understood by Einstein field equations, Prigogine et al
\cite{I.Prigogine1989} studied the evolution of the universe after
introducing particle creation mechanism in Einstein's field
equations by changing the usual balance equation for the number
density of particles.

In cosmological dynamics, the only dissipative phenomenon in the
homogeneous and isotropic flat
Friedmann-Lema{\^\i}tre-Robertson-Walker (FLRW)  model may be in
the form of bulk viscous pressure either due to coupling of
different components of the cosmic substratum
\cite{S.Weinberg1971, N.Straumann1976, M.A.Schweizer1982,
N.Udey1982, W.ZimdahlMNRAS1996} or due to non-conservation of
(quantum) particle number. Thus for an open thermodynamical system
where the number of fluid particles are not preserved
($N^{\mu}_{;\mu}\neq0$) \cite{Ya B.Zel'dovich1970, G.L.Murphy1973,
B.L.Hu1982}, the particle conservation equation gets modified as
\begin{equation}\label{particleconservation}
 N^{\mu}_{;\mu}\equiv n_{,\mu}u^{\mu}+\Theta n=n\Gamma\Longleftrightarrow N_{,\mu}u^{\mu}=\Gamma N,~~{\it i.e.},~~ \dot{N}=\Gamma N
 \end{equation}

This equation is also known as the balance equation for the particle flux. Also the implied relation states that the rate of change of total particle number is proportional
 to the total number of particles. Here, $\Gamma$ stands for the rate of change of particle number in a comoving volume $V$ containing $N$ number of particles, $N^{^{\mu}}=nu^{\mu}$,
the particle flow vector, $u^{\mu}$ is the four velocity vector,
$n=N/V$ is the particle number density and $\Theta=u^{\mu}_{;\mu}$
is the fluid expansion. The quantity $\Gamma$ is unknown in nature
but the validity of second law of thermodynamics implies the
positivity of $\Gamma$. In the present work, dissipative effect
due to the second alternative is chosen. However, for simplicity,
adiabatic ({\it i.e.}, isentropic) production
\cite{I.Prigogine1989, M.O.Calvao1992}  of perfect fluid particles
is considered and as a result viscous pressure obeys a linear
relationship with particle production rate.

The particle creation scenario can successfully describe the
accelerated expansion model of universe without introducing DE.
Also, many interesting results with this mechanism, such as, a
possibility of future deceleration has been proposed in
\cite{S.Pan2015, S.Chakraborty2014a}, consequently, an existence
of an emergent universe has been shown in
\cite{S.Chakraborty2014b,Dutta:2015fha} and subsequently, the
complete cosmic scenario has been reported in
\cite{S.Chakraborty2014c}. Further, in the framework of particle
creation mechanism universe evolves from big bang scenario to late
time de Sitter phase in \cite{Jaume de Haro2015} and accelerated
expansion of universe at early and present times are reported in
\cite{S.Pan2016}. Furthermore, the possibility of a phantom
universe without invoking any phantom fields has recently been
realized in the similar context \cite{RC.Nunes2015, RC.Nunes2016}.
So it is worth studying interacting DE models from particle
creation mechanism.

In the present work, considering our universe as an open
thermodynamical system in the framework of flat FLRW spacetime, an
interacting dynamics between dark energy and dark matter has been
proposed where the dark matter particles are assumed to be created
from the gravitational field. This is achieved by rewriting the
Friedmann equation and Raychaudhuri equation in the context of
matter creation mechanism and assuming the particle production
rate to be proportional to the Hubble parameter and is uniform
throughout the universe. The main scope of this work is to analyze
the cosmological dynamics of interacting  DE models in the
framework of adiabatic particle creation using dynamical system
techniques. Dynamical system tools have been extensively used to
study the asymptotic behavior of  various cosmological models
where exact solutions of evolution equations cannot be obtained
(see for e.g.,
\cite{Boehmer:2011tp,Tamanini,Dutta:2015jaq,Carloni:2015jla,Dutta:2016dnt,
Dutta:2016bbs,Granda:2016etr,Dutta:2017kch}). We obtained some
interesting critical points which describe many interesting
results from the phase space analysis of linear interactions.
These include  the early matter dominated universe, the late time
DE dominated attractors in some parameter region, where DE is
associated with quintessence, cosmological constant, or phantom
field respectively.

The organization of the paper is as follows: In Sec.
\ref{basic-equations}, we present the basic equations of the
present particle creation model and the evolution equations are
transformed to an autonomous system by suitable transformation of
the dynamical variables. In Sec. \ref{critical_points-parameters},
 critical points are shown for various choices for the interaction term and the cosmological parameters have been evaluated. Sec. \ref{phase-space}
 shows the phase space analysis and stability criteria for the critical points. In Sec. \ref{cosmological-implications},
  cosmological implication of critical points for several interaction models are given. The paper ends with a short discussion in sec. \ref{short-discussion}.\\

\section{The Basic Equations in particle creation and autonomous system}
\label{basic-equations}

In accordance  with inflation and cosmic microwave background
radiation, the universe is well described by the spatially flat
FLRW space-time

\begin{equation}\label{metric}
ds^2 = -dt^2+ a^2 (t) \Bigl(dr^2+ r^{2} d\Omega^2\Bigr),
\end{equation}
where $a (t)$ is the scale factor of the universe and the
spherical line element $d\Omega^{2}=d\theta^{2}+\sin^{2}\theta
d\phi^{2}$ is the metric on the unit 2-sphere. For the co-moving
observer, $u^{\mu}=\delta^{\mu}_{t}$, is the velocity vector so
that $u^{\mu}u_{\mu}=-1$, accompanied with the line element
(\ref{metric}) and consequently the fluid expansion ($\Theta$)
will become, $\Theta=3H$, where $H$ is the Hubble parameter.
Hence, the particle conservation equation
(\ref{particleconservation}) is reduced to

\begin{equation}\label{revisedparticleconservation}
 N^{\mu}_{;\mu}\equiv n_{,\mu}u^{\mu}+3H n=n\Gamma,
\end{equation}
for the present open thermodynamical model.
Further, using the above conservation equation (\ref{revisedparticleconservation}), the Gibb's relation \cite{I.Prigogine1989, W.Zindahl1996, W.Zindahl2000}

\begin{equation}\label{Gibbsrelation}
T ds=d\left(\frac{\rho}{n} \right)+p d\left( \frac{1}{n}\right),
\end{equation}
gives the variation of entropy per particle by the relation \cite{I.Prigogine1989, Jaume de Haro2015}

\begin{equation}\label{entropyrelation}
n T\dot{s}=\dot{\rho}+3H\left( 1-\frac{\Gamma}{3H}\right)(\rho+p),
\end{equation}
where $T$ represents the fluid temperature, $s$ is the entropy per
particle, {\it i.e.}, specific entropy (specific entropy of a
system is the entropy of the unit mass of the system), $\rho$ is
the total energy density and $p$ denotes the total thermodynamic
pressure. We consider our thermodynamical system to be ideal one,
{\it i.e.}, isentropic (or adiabatic) (see, for instance
\cite{M.O.Calvao1992, J.D.Barrow1990}) and consequently we have
the production of perfect fluid particles with constant entropy
({\it i.e.}, $\dot{s}=0$). However, there is entropy production
due to the enlargement of the phase space of the system since
number of perfect fluid particles increases. Hence, from equation
(\ref{entropyrelation}) one can obtain the conservation equation
as

\begin{equation}
\dot{\rho}+3H(\rho+p)=\Gamma (\rho+p),~~ \label{conservation-relation}
\end{equation}
which however can also be written as \cite{S.Pan2016}

\begin{equation}
\dot{\rho}+3H(\rho+p+p_{c})=0.~~ \label{Mod-consv-Eqn}
\end{equation}
So, comparing  equations (\ref{conservation-relation}), and (\ref{Mod-consv-Eqn}), one can easily obtain the creation pressure as
 follows \cite{J.A.S.Lima2012, J.P.Mimoso2013, Jaume de Haro2015, S.Chakraborty2014c}

\begin{align}
p_{c}=-\frac{\Gamma}{3H}(\rho+p), \label{Cp}
 \end{align}
where $p_{c}$ is termed as the creation pressure, and $\Gamma$ is the particle production rate (number of particles created per unit time),
 assumed to be uniform throughout the universe. Now, considering the main constituents of our universe as dark matter (DM) in the form of dust
  having energy density $\rho_{m}$  and the dark energy
   fluid having equation state   $\omega_{d}=p_{d}/\rho_{d}$ (where $\rho_{d}$, $p_{d}$ are respectively the energy density and the thermodynamic
   pressure of the cosmic fluid). The  Friedmann and Raychaudhuri equations ($8\pi G=c=1$) can now be written as:

\begin{align}
H^{2} &=\frac{1}{3}(\rho_{m}+\rho_{d}),\label{friedmann}~~~~~~~~~~~~~~~~~~~~~~~(\mbox{Friedmann's equation})\\
\dot{H} &=-\frac{1}{2}\Bigl(\rho_{m}+\rho_{d}+p_{d}+p_{c}\Bigr),\label{Raychaudhuri}~~~~~~(\mbox{Raychaudhuri's equation})
 \end{align}
where $H= \dot{a}/a$ is the Hubble parameter and an `overdot'
represents the differentiation with respect to cosmic time `t'. If
we assume that the  created particles are described as
pressure-less dark matter (DM) in the thermodynamically open model
of the universe under the adiabatic condition, then  the  creation
pressure $p_{c}$ in (\ref{Cp}) becomes \cite{RC.Nunes2015}

$$ p_{c}=-\frac{\Gamma}{3H}(\rho_{m}).$$

In standard cosmology, the dynamic interactions between the homogeneously distributed DE in the universe and the DM component
(clumping around the ordinary particles) are extremely weak or even it is negligible. As a result,
the energy conservation equations for the two matter components are
\begin{align}
\dot{\rho}_m +3H(\rho_{m}+p_{c})=0
\end{align}
or, using the above expression for $p_{c}$ we have
\begin{align}
\dot{\rho}_m +3H\rho_{m}=\Gamma\rho_{m}
\end{align}

and
\begin{align}
\dot{\rho}_{d} +3H(\rho_{d}+p_{d}) &= 0.\label{fc}
\end{align}

In order to alleviate the cosmological coincidence problem, it has
been found that a non-gravitational interaction between these dark
sectors could be a viable alternative. So the interacting DM and
DE models of the universe are becoming of great interest and are
widely used in the literature \cite{T.Harko2013}. Thus the above
energy conservation equations are modified as :

\begin{align}
\dot{\rho}_m +3H\rho_{m}\Bigl(1-\frac{\Gamma}{3H}\Bigr) & = -Q,\label{mc}\\
\dot{\rho}_{d} +3H\Bigl(\rho_{d}+p_{d}\Bigr) &= Q,\label{fc}
\end{align}

where $Q$ indicates the rate of energy exchange between the dark sectors.

In particular, $Q>0 $ indicates conversion of DM into DE, while
$Q<0$ represents the vice-versa. A complete study of the
interaction of dynamical vacuum
 energy with matter can be found in the references \cite{Sola6,Sola7} (for an extension see also the Refs. \cite{Sola8,Sola9}). Further, it should be noted
   that the running vacuum models \cite{Sola10,Sola11} give the overall fit to the observational data better than the $\Lambda$CDM.  These studies are based
    on the general expectations of the effective action of quantum field theory ($QFT$) in curved space time  and provide an interaction of dynamical vacuum and matter\cite{Sola12}.

In the present work, we describe the background dynamics with different  interactions as
 $(i)~~ Q\propto H\rho_{m}~~$ \cite{C.G.Bohmer2008, Xi-ming Chen2009},
 $(ii)~~Q\propto H\rho_{d}~~$ \cite{J.-H2008, D. Pavon2009},~~ and~~
$(iii)~~ Q\propto H(\rho_{m}+\rho_{d})$ \cite{R.Garcia-Salcedo2012, Andre A.Costa2014},
$(iv)~~Q\propto \frac{\rho_{m}\rho_{d}}{H}$ \cite{S.Pan2014}, ~~and
$(v)~~Q\propto \rho_{m}$ \cite{C.G.Bohmer2008}.\\

One may note that in the above interactions, the dimensionless parameters (proportionality constants) should not
 be chosen in an ad hoc manner. These parameters can actually be fitted to the overall observational data and one
 finds that they are typically of order between $10^{-3}$ to $10^{-2}$ \cite{Sola9,Sola10,Sola11,Sola13} depending on the normalization of
  the parameters involved. Further, in these references the justifications of such small values of these parameters are shown from two perspectives,
   namely theoretically these coefficients represent the beta-function of the running vacuum energy \cite{Sola11,Sola13} and hence are expected
   to be very small. Also, experimentally, as the fitted values of these coefficients to the recent SNIa $+$ BAO $+$ LSS $+$ BBN $+$ CMB data
   (in which WMAP9, Planck-13, 15 data are taken into account) \cite{Ade:2015xua} are found to be of the same order as the theoretically expected values.
    Further, if we compare our equation (\ref{mc}) with equation (4) of the $Q_{m}$ model in Ref. \cite{Sola13} and choosing $\Gamma=\Gamma_{0}H $ ($\Gamma_{0}$, a constant),
     we see that effective interaction term will be $(\Gamma_{0}-\alpha_{m})H \rho_{m}$ in the first case. Thus comparing equation (7) of Ref. \cite{Sola13} we have
\begin{align}
\Gamma_{0}-\alpha_{m}=3\nu_{dm}\label{effint}.
\end{align}

Moreover, recent observationally estimated values of the parameters $\nu_{dm}$ and $\nu_{\Lambda}$ (in equations (7) and (8) of Ref. \cite{Sola13}) similar to our interaction models 1 and 2 are given by (see table II of Ref. \cite{Sola13})
\begin{eqnarray}
\nu_{dm}=0.00618\pm 0.00159,\nonumber\\
\nu_{\Lambda}=0.01890\pm 0.00744.\nonumber
\end{eqnarray}
Thus, $\Gamma_{0}$ and $\alpha_{m}$ are  not arbitrary, their difference has an observational estimate. Note that as $\nu_{dm}$ is positive so from equation (\ref{effint}),
 $\Gamma_{0}$ is always greater than $\alpha_{m}$ and the effective interaction term has the same sign convention as in Ref. \cite{Sola13}.\\

Due to complicated non-linear forms in the evolution equations (\ref{friedmann}), (\ref{Raychaudhuri}), (\ref{mc}) and (\ref{fc}) we convert these evolution equations to an autonomous system of first order differential equations.
 To do this we consider the following dimensionless variables \cite{M.Khurshudyan2015}
\begin{align}\label{variables}
x=\frac{\rho_{d}}{3H^{2}},~  y=\frac{p_{d}}{3H^{2}},
\end{align}
which are normalized over Hubble scale.

Then, the  autonomous system of ordinary differential equations obtained are:
\begin{eqnarray}
\frac{dx}{dN} &=& \frac{Q}{3H^{3}}-(1-x)\Bigl(3y+\Gamma_{0} x \Bigr),~ \label{de1}\\
\frac{dy}{dN} &= & \frac{Q y}{3x H^{3}}-(1-x)\left(\frac{3y^{2}}{x}+\Gamma_{0}y \right).~ \label{de2}
\end{eqnarray}
Here, the independent variable is chosen as the lapse time $ N=\ln a $, which is called the e-folding number and the particle production rate $\Gamma$ as a function of the Hubble parameter
\cite{S.Pan2015, S.Chakraborty2014c}  ($\Gamma$ has dimensions  $({\rm time})^{-1}$) is chosen as above $ \Gamma=\Gamma_{0}H $ ($\Gamma_{0}$ is a constant). The value of the parameter $\Gamma_{0}$ is assumed to be nonnegative as the creation of particles are only considered in this study.

Now, in terms of the new dimensionless quantities, cosmological parameters can be written as follows. For instance,
the energy density parameter for dark matter as
 \begin{align}
    \Omega_{m}=1-x,
\end{align}
and the energy density parameter for the dark energy as
\begin{align}
    \Omega_{d}=x.
\end{align}

It may be noted that in the case of  non-interacting DE models,
the energy density is usually considered to be non-negative.
However, in this case of interacting DE models, the energy density
can be taken to be negative \cite{Quartin:2008px}. This would
imply that there is no constrain for dimensionless variables,
making the phase space that is analyzed here to be not compact.

So, there might be a possibility of critical points at infinity.
In general, the analysis of fixed points at infinity is done by
compactifying the phase space using Poincare compactification.
However, from a phenomenological point of view in this present
work, we shall only determine the dynamics in the neighborhood of
finite fixed points. This is enough, since our aim is to find
physically viable solutions, namely trajectories connecting DM to
DE domination.

Also, the equation of state parameter for the DE can be expressed
as
\begin{align}\label{de-eos}
\omega_{d}=\frac{p_{d}}{\rho_{d}}=\frac{y}{x}~,
\end{align}
and the effective equation of state parameter will be of the form

\begin{align}\label{eff-eos}
\omega_{\rm eff} & = y-\frac{\Gamma_{0}}{3}\Bigl(1-x\Bigr)~.
\end{align}

 Moreover, we have the evolution equation of the Hubble function as
 \begin{equation}\label{Hubble-fn}
 \frac{1}{H}\frac{dH}{dN} =-\frac{3}{2}\Bigl(1+y-\frac{\Gamma_{0}}{3}(1-x)\Bigr)~.
\end{equation}

We now determine the critical points of the above autonomous system for different choices of $Q$ and then we perturb the equations
 up to first order about the critical points, in order to determine their stability.

\section{Critical points of autonomous system (\ref{de1})--(\ref{de2}) for various choices of interaction term and the cosmological parameters:}
\label{critical_points-parameters}

In this section, we discuss  the existence of the critical points and the corresponding physical parameters for various interaction models. These are presented in detail in tabular form.

\subsection{Interaction Model 1:}
 First, we choose the interaction as
 \begin{equation}\label{1stInteraction}
 Q=\alpha_{m}H\rho_{m}~,
 \end{equation}
where the coupling parameter $\alpha_{m}$ is a dimensionless constant. The indefiniteness in the sign of $\alpha_{m}$ indicates that
the energy transfer takes place in the either direction - DE or DM. This interaction is well motivated due to mathematical simplicity
 as the dimensions of the autonomous system (\ref{de1})-(\ref{de2}) remain same because $H$ parameter can be eliminated from the equations. Now, using this interaction
  in the system (\ref{de1})-(\ref{de2}), the autonomous system for this interaction model will be
 \begin{eqnarray}
\frac{dx}{dN} &=& (-1+x)(\Gamma_0 x-\alpha_m+3y),\label{Int1de1}\\
\frac{dy}{dN} &= & \frac{y}{x} (-1+x)(\Gamma_0 x-\alpha_m+3 y).\label{Int1de2}
\end{eqnarray}

The critical points for this system (\ref{Int1de1})-(\ref{Int1de2})  are the following:\\
\begin{itemize}
\item Set of critical points: $ A_{1}  = ( 1, y_c )$, where $y_c$ takes any real value.

\item Critical Point : $ B_{1}= ( \frac{\alpha_{m}}{\Gamma_{0}}, 0  ) $.

\item Critical point  :$ C_{1} = (1,0) $.

\item Critical point  :$ D_{1} = (1,-1) $.

\item Set of critical points  :$ E_{1} = (x_c, \frac{\alpha_{m}}{3}-\frac{\Gamma_{0} x_c}{3}) $.

\end{itemize}

The existence of critical points and their cosmological parameters are displayed in the table \ref{modelQ1}. It is observed
that point $B_1$ is a point on the set $E_1$, points $C_1$ and $D_1$ are points on the set $A_1$. So, in the next section,
 we shall analyze only the stability of sets $A_1$ and $E_1$. However, critical points $B_1$, $C_1$ and $D_1$ show some interesting
 cosmological feature which we shall discuss in Sec. \ref{cosmological-implications}.  \\

\begin{table}[tbp] \centering
\caption{The existence of critical points and the corresponding physical parameters for the interaction model $Q_{1}=\alpha_{m}H\rho_{m}$.}
\begin{tabular}
[c]{cccccc}\hline\hline
\textbf{Critical Points} & \textbf{Existence} &$\mathbf{\omega_{d}}$ &
$\mathbf{\omega_{eff}}$ & $\mathbf{\Omega_{m}}$ & $\mathbf{\Omega_{d}}$ \\\hline
$A_{1}:(1,y_c)  $  & Always &$ y_c $  & $ y_c $   &    $ 0 $& $ 1 $   \\
$B_{1}:\left(\frac{\alpha_{m}}{\Gamma_{0}},0\right)  $ & $\Gamma_{0}\neq 0$ & $ 0  $ &
$-\frac{\Gamma_{0}}{3}(1-\frac{\alpha_{m}}{\Gamma_{0}})$ & $ 1-\frac{\alpha_{m}}{\Gamma_{0}} $ & $ \frac{\alpha_{m}}{\Gamma_{0}} $ \\
$C_{1}:(1,0)  $ & Always & $0$ & $0$ &
$0$ & $1$ \\
$D_{1}:(1,-1)  $  & Always &$ -1 $  & $ -1 $   &    $ 0 $& $ 1 $   \\
$E_{1}:\left(x_c,~ \frac{\alpha_{m}}{3}-\frac{\Gamma_{0}x_c}{3}\right)  $  & Always &$ \frac{\alpha_{m}-\Gamma_{0}x_c}{3x_c} $  & $ \frac{\alpha_{m}}{3}-\frac{\Gamma_{0}}{3} $   &    $ 1-x_c $& $ x_c $  \\
\hline\hline
\end{tabular}
\label{modelQ1}
\end{table}%
%

\subsection{Interaction Model 2:}

We consider another choice of interaction as
\begin{equation}\label{2ndInteraction}
Q=\alpha_{d}H\rho_{d},
\end{equation}
where $\alpha_{d}$ is the coupling parameter. Using this
interaction in the system (\ref{de1})-(\ref{de2}), we have the
following autonomous system:
\begin{eqnarray}
\frac{dx}{dN} &=& (\Gamma_0 x+3 y) (x-1)+\alpha_d x,\label{Int2de1}\\
\frac{dy}{dN} &= &\frac{y}{x} \Big((\Gamma_0 x+3 y) (x-1)+\alpha_d x \Big).\label{Int2de2}
\end{eqnarray}

The autonomous system (\ref{Int2de1})-(\ref{Int2de2}) admits the following critical points:\\

\begin{itemize}
\item Critical Point : $ A_{2}= \Bigl(\frac{\Gamma_{0}-\alpha_{d}}{\Gamma_{0}}, 0 \Bigr )$.
\item Set of critical points  : $ B_{2}= \Bigl( x_c, \frac{x_c (\Gamma_{0} x_c-\Gamma_{0}+\alpha_{d})}{3(1-x_c)} \Bigr).$
\end{itemize}

The existence criteria  and the cosmological parameter related to the critical points are shown in the table \ref{modelQ2}.  It is again noted that point $A_2$ is a point on a set $B_2$. So, in the next section we shall analyze only the stability of set $B_2$. However, depending on the choice of coupling parameter $\alpha_d$, critical point $A_2$ shows some interesting cosmological features, which we shall discuss in Sec. \ref{cosmological-implications}. \\

\begin{table}[tbp] \centering
\caption{The existence of critical points and the corresponding physical parameters for the interaction model $Q_{2}=\alpha_{d}H\rho_{d}$.}
\begin{tabular}
[c]{cccccc}\hline\hline
\textbf{Critical Points} & \textbf{Existence} &$\mathbf{\omega_{d}}$ &
$\mathbf{\omega_{eff}}$ & $\mathbf{\Omega_{m}}$ & $\mathbf\Omega_{d}$\\\hline
$A_{2}:\left(\frac{\Gamma_{0}-\alpha_{d}}{\Gamma_{0}},0\right)  $ &$\Gamma_{0} \neq 0
$& $ 0 $& $ -\frac{\alpha_{d}}{3}$ & $ \frac{\alpha_{d}}{\Gamma_{0}}$ & $1-\frac{\alpha_{d}}{\Gamma_{0}}$ \\
$B_{2}:\left(x_c,\frac{x_c(\Gamma_{0}x_c-\Gamma_{0}+\alpha_{d})}{3(1-x_c)}\right)  $&$x_c \neq 1$ &$\frac{(\Gamma_{0}x_c-\Gamma_{0}+\alpha_{d})}{3(1-x_c)}
$ &$ \frac{(\Gamma_{0}x_c-\Gamma_{0}+\alpha_{d}x_c)}{3(1-x_c)}$ & $1-x_c$ & $x_c$ \\
\hline\hline
\end{tabular}
\label{modelQ2}
\end{table}%
%

\subsection{Interaction Model 3:}

Now, we consider the linear interaction as

\begin{equation}\label{3rdInteraction}
Q=\alpha H(\rho_{m}+\rho_{d}).
\end{equation}
For this interaction model, the system (\ref{de1})-(\ref{de2})
will take the form
\begin{eqnarray}
\frac{dx}{dN} &=& \alpha+(\Gamma_0 x+3y) (x-1),\label{Int3de1}\\
\frac{dy}{dN} &= & \frac{y}{x}\left(\alpha+(\Gamma_0 x+3y) (x-1)\right).\label{Int3de2}
\end{eqnarray}

The critical points are:

\begin{itemize}
\item  Critical Point : $ A_{3}: \left(\frac{1}{2}(1+\sqrt{1-\frac{4\alpha}{\Gamma_{0}}}),~ 0\right) $, where
$\Gamma_{0}\geq 4\alpha$.

\item Critical Point  : $ B_{3}: \left(\frac{1}{2}(1-\sqrt{1-\frac{4\alpha}{\Gamma_{0}}}),~ 0\right) $, where
$\Gamma_{0}\geq 4\alpha$.

\item Set of critical points :$ C_{3}: \left(x_c,~ \frac{\Gamma_{0}x_c^{2}-\Gamma_{0}x_c+\alpha}{3(1-x_c)}\right) $.

\end{itemize}

The cosmological parameters related to the critical points are shown in tabular form in table \ref{modelQ3}.  It is again noted that points $A_3$, $B_3$ are points on a set $C_3$. So, in the next section we shall analyze only the stability of set $C_3$. \\

\begin{table}[tbp] \centering
\caption{The existence of critical points and the corresponding physical parameters for the interaction model $Q_{3}=\alpha H(\rho_{m}+\rho_{d})$}
\begin{tabular}
[c]{cccccc}\hline\hline
\textbf{Critical Points} & \textbf{Existence} &$\mathbf{\omega_{d}}$ &
$\mathbf{\omega_{\rm eff}}$ & $\mathbf{\Omega_{m}}$ & $\mathbf\Omega_{d}$ \\\hline
$A_{3}:\left(\frac{1}{2}(1+\sqrt{1-\frac{4\alpha}{\Gamma_{0}}}),~ 0\right)  $ & $\frac{\alpha}{\Gamma_0}<\frac{1}{4}
$&$0$ & $\frac{\Gamma_{0}}{6}(-1+\sqrt{1-\frac{4\alpha}{\Gamma_{0}}}) $ & $\frac{1}{2}(1-\sqrt{1-\frac{4\alpha}{\Gamma_{0}}})$ & $ \frac{1}{2}(1+\sqrt{1-\frac{4\alpha}{\Gamma_{0}}})$\\
$B_{3}:\left(\frac{1}{2}(1-\sqrt{1-\frac{4\alpha}{\Gamma_{0}}}),~ 0\right)  $ &  $\frac{\alpha}{\Gamma_0}<\frac{1}{4}$&$0$ & $-\frac{\Gamma_{0}}{6}(1+\sqrt{1-\frac{4\alpha}{\Gamma_{0}}}) $ & $\frac{1}{2}(1+\sqrt{1-\frac{4\alpha}{\Gamma_{0}}})$ & $ \frac{1}{2}(1-\sqrt{1-\frac{4\alpha}{\Gamma_{0}}})$\\
$C_{3}:\left(x_c,~ \frac{\Gamma_{0}x_c^{2}-\Gamma_{0}x_c+\alpha}{3(1-x_c)}\right)  $  &$x_c \neq 1$& $\frac{\Gamma_{0}x_c^{2}-\Gamma_{0}x_c+\alpha}{3x_c(1-x_c)}$ &$\frac{\Gamma_{0}x_c-\Gamma_{0}+\alpha}{3(1-x_c)}
$ & $ 1-x_c$ & $x_c $ \\\hline\hline
\end{tabular}
\label{modelQ3}
\end{table}%

\subsection{Interaction Model 4:}

Considering the  nonlinear interaction

\begin{equation}\label{4thInteraction}
Q=\frac{\beta}{H}\rho_{m}\rho_{d}
\end{equation}
in (\ref{de1})-(\ref{de2}), the autonomous system will be in the
form
\begin{eqnarray}
\frac{dx}{dN} &=&(x-1)(\Gamma_0 x-3\beta x+3y),\label{Int4de1}\\
\frac{dy}{dN} &= & \frac{y}{x}(x-1)(\Gamma_0 x-3\beta x+3y),\label{Int4de2}
\end{eqnarray}

and the following are the critical points for this case:
\begin{itemize}
\item Set of critical points  : $ A_{4}= ( 1,~ y_c ) $.
\item Critical Point  : $ B_{4}= ( 1, 0 ) $.
\item Critical Point  : $ C_{4}= ( 1, -1 ) $.
\item Set of critical points : $ D_{4}= \left( x_c,~ \beta x_c-\frac{\Gamma_{0} x_c}{3} \right)$.

\end{itemize}

The condition for existence of critical points and the corresponding physical parameters are presented in table \ref{modelQ4}.
 It is again noted that points $B_4$, $C_4$ are points on the set $A_4$. So, in the next section we shall analyze only the stability of sets $A_4$ and $D_4$.

\begin{table}[tbp] \centering
\caption{The existence of critical points and the corresponding physical parameters for the interaction model $Q_{4}=\frac{\beta}{H}\rho_{m}\rho_{d}$}
\begin{tabular}
[c]{cccccc}\hline\hline
\textbf{Critical Points} & \textbf{Existence} &$\mathbf{\omega_{d}}$ &
$\mathbf{\omega_{eff}}$ & $\mathbf{\Omega_{m}}$ & $\mathbf\Omega_{d}$ \\\hline
$A_{4}:(1,y_c)  $ & Always & $ y_c $  &  $ y_c $  & 0 & 1 \\
$B_{4}:(1,0)  $ & Always &   0     &    0     &     0     &     1    \\
$C_{4}:(1,-1)  $ & Always & $ -1 $  &  $ -1 $  & 0 & 1\\
$D_{4}:\left(x_c,~ \beta x_c-\frac{\Gamma_{0}x_c}{3}\right)  $ & Always & $ \beta-\frac{\Gamma_{0}}{3}$ & $\beta x_c-\frac{\Gamma_{0}}{3}$  & $1-x_c$   & $ x_c $
\\\hline\hline
\end{tabular}
\label{modelQ4}
\end{table}%
%

\subsection{Interaction Model 5:}

We are now going to discuss another type of interaction in the dark sectors which is completely based on
 the local properties of the universe and hence it is different from the other interaction models discussed
 in the previous subsections. Here,  we replace the non local transfer rate (discussed in the previous sub-sections)
 by the local rate $\eta$, and the interaction \cite{C.G.Bohmer2008} has the following form

\begin{equation}\label{Int5}
Q=\eta \rho_{m},
\end{equation}
where the coefficient $\eta$ related to the local rate is assumed
to be constant. When $\eta>0$ (i.e., $Q>0$), the energy decays
from DM to DE which reveals the possibility of vanishing DE field
in the primordial universe, so that DE condenses as a result of
slow decay of DM.  This interaction has been studied in
\cite{Xi-ming Chen2009} describing the phase space analysis when
the dark energy equation of state has the phantom behavior.
Moreover, for $\eta>0$ the interaction is used for describing the
decay of DM into radiation \cite{R.Cen2001}, the decay of a
curvaton field into radiation \cite{K.A.Malik2003} and also a
model with the decay
  of superheavy DM particles into a quintessence scalar field \cite{H.Ziaeepour2004}.
   On the other hand, for $\eta<0$ the energy is transferred in the reverse way.
    Further, in order to  close the dynamical system (\ref{de1})-(\ref{de2}),
     one has to introduce the new variable $v$ given by

\begin{equation}
v=\frac{H_{0}}{H+H_{0}},
\end{equation}
where $H_{0}$ is constant and hence $0\leq v\leq1$.\\
Introducing a dimensionless coupling constant
\begin{equation}
\gamma=\frac{\eta}{H_{0}},
\end{equation}
 the autonomous system of equations can be written as
\begin{eqnarray}
\frac{dx}{dN} &=& \frac{(-1+x)}{(-1+v)}\left((\Gamma_0 x+3y)(v-1)+\gamma\,v \right),\label{Int5de1}\\
\frac{dy}{dN} &= &  \frac{y(-1+x)}{x(-1+v)}\left((\Gamma_0 x+3y)(v-1)+\gamma\,v \right),\label{Int5de2}\\
\frac{dv}{dN} &= & \frac{1}{2}v(v-1)\Bigl(3+3y-\Gamma_{0}(1-x)\Bigr).\label{Int5de3}
\end{eqnarray}
The critical points for this system are the following:

\begin{itemize}
\item  Set of critical points  : $ A_{5}= ( 1,~ y_c, 0 ) $. \item
Set of critical points : $ B_{5}=(1,-1, v ) $. \item Set of
critical points : $ C_{5}= \left( x_c,~ -\frac{\Gamma_{0}
x_c}{3},~ 0 \right) $. \item  Set of critical points : $ D_{5}=
\Big( x_c,~ -1+\frac{\Gamma_{0}}{3}(1-x_c),~
\frac{-3+\Gamma_{0}}{-3+\Gamma_{0}+\gamma} \Big) $.
\end{itemize}

The set of critical points and the corresponding cosmological parameters are presented in Table \ref{modelQ5}. We note here that sets $A_5$ and $B_5$ have common point $(1,-1,0)$.\\

\begin{table}[tbp] \centering
\caption{The existence of critical points and the corresponding physical parameters for the interaction model $Q_{5}=\eta\rho_{m}$}%
\begin{tabular}
[c]{ccccccc}\hline\hline
\textbf{Critical Points} & \textbf{Existence} &$\mathbf{\omega_{d}}$ &
$\mathbf{\omega_{eff}}$ & $\mathbf{\Omega_{m}}$ & $\mathbf\Omega_{d}$\\\hline
$A_{5}:(1,y_c,0)  $ &Always
 &  $ y_c $  & $y_c$  & 0 & 1 \\
$B_{5}:(1,-1,v)  $ & $0\leq v\leq1$ &  $ -1$ & $-1$ & 0 & 1 \\
$C_{5}:\left(x_c,-\frac{\Gamma_{0}x_c}{3},0\right)  $ & Always
& $-\frac{\Gamma_{0}}{3}$ & $-\frac{\Gamma_{0}}{3}$  & $ 1-x_c $   &    $x_c$\\
$ D_{5}:\left(x_c,-1+\frac{\Gamma_{0}}{3}(1-x_c),\frac{-3+\Gamma_{0}}{-3+\Gamma_{0}+\gamma}\right) $ & $\Gamma_0+\gamma \neq 3$  &$-\frac{1}{3x_c}(3-\Gamma_{0}+\Gamma_{0}x_c)$     &   $ -1$     &  $1-x_c$     &  $x_c$   \\
\hline\hline
\end{tabular}
\label{modelQ5}
\end{table}%
%



\section{Phase space analysis and Stability criteria of critical points }
\label{phase-space}

\begin{table}
 \caption{ Table shows the eigenvalues of the critical points for different interaction model}  \label{evalues2D}
 \begin{tabular}{|c|c|c|c|c|c|c|c|c|c|}
 \hline
   $\mbox{Interaction}$ & $\mbox{Critical points} $ & $\lambda_{1} $ & $ \lambda_{2} $ \\\hline
   $ 1.Q=\alpha_{m}H\rho_{m}$   &   $ A_{1}:(1,y_c)  $  & 0  & $ 3y_c+\Gamma_{0}-\alpha_{m} $      \\\hline

   $ ,, $ & $B_{1}:(\frac{\alpha_{m}}{\Gamma_{0}},0)$  & 0  & $-\Gamma_{0}+\alpha_{m} $   \\\hline
   $ ,, $ & $C_{1}:(1,0)$  & 0  &$\Gamma_{0}-\alpha_{m}$      \\\hline
   $ ,, $   &   $ D_{1}:(1,-1)  $  & 0  & $ -3+\Gamma_{0}-\alpha_{m} $      \\\hline
   $ ,, $   &   $ E_{1}:(x_c,\frac{\alpha_{m}}{3}-\frac{\Gamma_{0}x_c}{3})  $  & 0  & $ \frac{\alpha_{m}(-1+x_c)}{x_c} $      \\\hline
   $2.Q=\alpha_{d}H\rho_{d} $ & $ A_{2}:\left(\frac{\Gamma_{0}-\alpha_{d}}{\Gamma_{0}},0\right) $ & 0 &   $\Gamma_{0}-\alpha_{d}$   \\\hline
   $ ,, $ & $ B_{2}:\left(x_c,\frac{x_c(\Gamma_{0}x_c-\Gamma_{0}+\alpha_{d})}{3(1-x_c)}\right) $ & 0 & $\frac{\alpha_{d}x_c}{1-x_c}$ \\\hline
   $3.Q=\alpha H\Bigl(\rho_{m}+\rho_{d}\Bigr) $ & $ A_{3}:\left(\frac{1}{2}(1+\sqrt{1-\frac{4\alpha}{\Gamma_{0}}}),0\right) $ & 0 & $\Gamma_{0}\sqrt{1-\frac{4\alpha}{\Gamma_{0}}}$ \\\hline
   $ ,, $ & $ B_{3}:\left(\frac{1}{2}(1-\sqrt{1-\frac{4\alpha}{\Gamma_{0}}}),0\right) $ & 0 & $-\Gamma_{0}\sqrt{1-\frac{4\alpha}{\Gamma_{0}}}$ \\\hline
   $ ,, $ & $ C_{3}:\left(x_c,\frac{\Gamma_{0}x_c^{2}-\Gamma_{0}x_c+\alpha}{3(1-x_c)}\right)  $ & 0 & $\frac{\alpha(2x_c-1)}{x_c(1-x_c)}$ \\\hline
   $ 4.Q=\frac{\beta}{H}\rho_{m}\rho_{d} $ & $A_{4}:(1,y_c)$  & 0& $ 3y_c+\Gamma_{0}-3\beta$  \\\hline
   $ ,, $ & $B_{4}:(1,0)$ & 0 & $\Gamma_{0}-3\beta$  \\\hline

   $ ,, $ & $C_{4}:(1,-1)$  & 0& $ -3+\Gamma_{0}-3\beta$  \\\hline
   $ ,, $   &   $ D_{4}:\left(x_c,\beta x_c-\frac{\Gamma_{0}x_c}{3}\right)  $  & 0    & 0    \\\hline

  \hline
   \end{tabular}
 \end{table}


\begin{table}
 \caption{ Table shows the eigenvalues of the critical points for the interaction model 5. $ Q=\gamma H_{0}\rho_{m} $.}\label{evalues3D}
 \begin{tabular}{|c|c|c|c|c|c|c|c|c|c|}
 \hline
   $ \mbox{Critical points}$ & $\lambda_{1} $ & $ \lambda_{2} $ & $ \lambda_{3} $ \\\hline
   $ A_{5}:(1,y_c,0) $ & 0  & $ \frac{3}{2}(1+y_c)$  & $ 3y_c+\Gamma_{0}$  \\\hline
   $ B_{5}:(1,-1,v)$& 0  & 0  & $ -3+\Gamma_{0}-\frac{\gamma v}{1-v} $     \\\hline
   $ C_{5}:(x_c,-\frac{\Gamma_{0}x_c}{3},0) $   &  0  &  0  & $\frac{1}{2}(3-\Gamma_{0})$     \\\hline
   $ D_{5}:\left(x_c,~ -1+\frac{\Gamma_{0}}{3}(1-x_c),~ \frac{-3+\Gamma_{0}}{-3+\Gamma_{0}+\gamma}\right)  $ & 0 & $ \frac{(x_c-1+\sqrt{1-x_c^{2}})(-3+\Gamma_{0})}{2x_c} $ & $-\frac{(-x_c+1+\sqrt{1-x_c^{2}})(-3+\Gamma_{0})}{2x_c}$   \\\hline
  \hline
 \end{tabular}
 \end{table}

We shall now discuss the phase space analysis of critical points and their stability by analyzing the eigenvalues of the linearized Jacobian matrix
evaluated at the critical points presented in Tables \ref{evalues2D} and \ref{evalues3D}. It can be seen from tables \ref{evalues2D} and \ref{evalues3D}
 that all critical points are actually non-isolated set of critical points. It is also noted that Eqn. (\ref{de2}) is $\frac{y}{x}$ times of Eqn. (\ref{de1}), but this does not imply that Eqn. (\ref{de2}) is obtained from Eqn. (\ref{de1}) as the equation of state parameter ($\omega_d=\frac{y}{x}$) is not a constant. As a result, all critical points obtained are non-isolated sets.  By definition, non-isolated set contains atleast one vanishing eigenvalue,
  so it is non-hyperbolic in nature \cite{Coley}. The type of non-isolated set with exactly one vanishing eigenvalue is called a normally hyperbolic set.
   Its stability condition is similar to the linear stability analysis and can be determined simply by looking from the signature of the remaining non-vanishing eigenvalues \cite{Coley}.
   In this work, all set of points are normally hyperbolic except set $B_5$, $C_5$
and $D_4$ (see Tables \ref{evalues2D} and \ref{evalues3D}).

\subsection{Interaction 1}
\begin{figure}
\centering
\subfigure[]{%
\includegraphics[width=6cm,height=4cm]{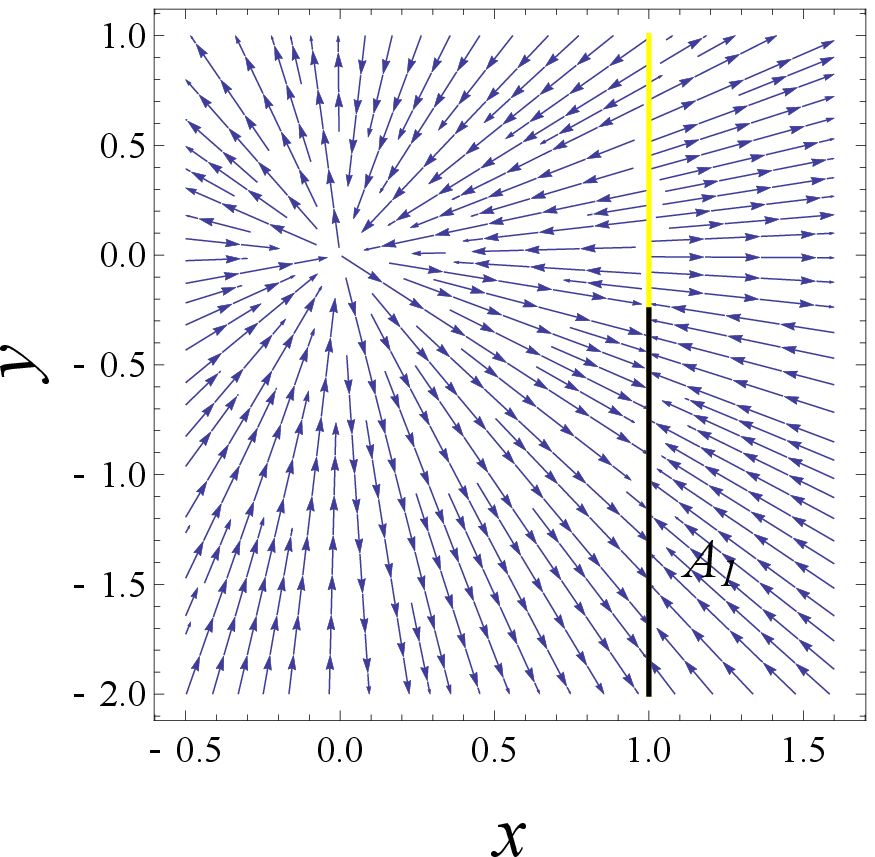}\label{fig:stable_A1}}
\qquad
\subfigure[]{%
\includegraphics[width=6cm,height=4cm]{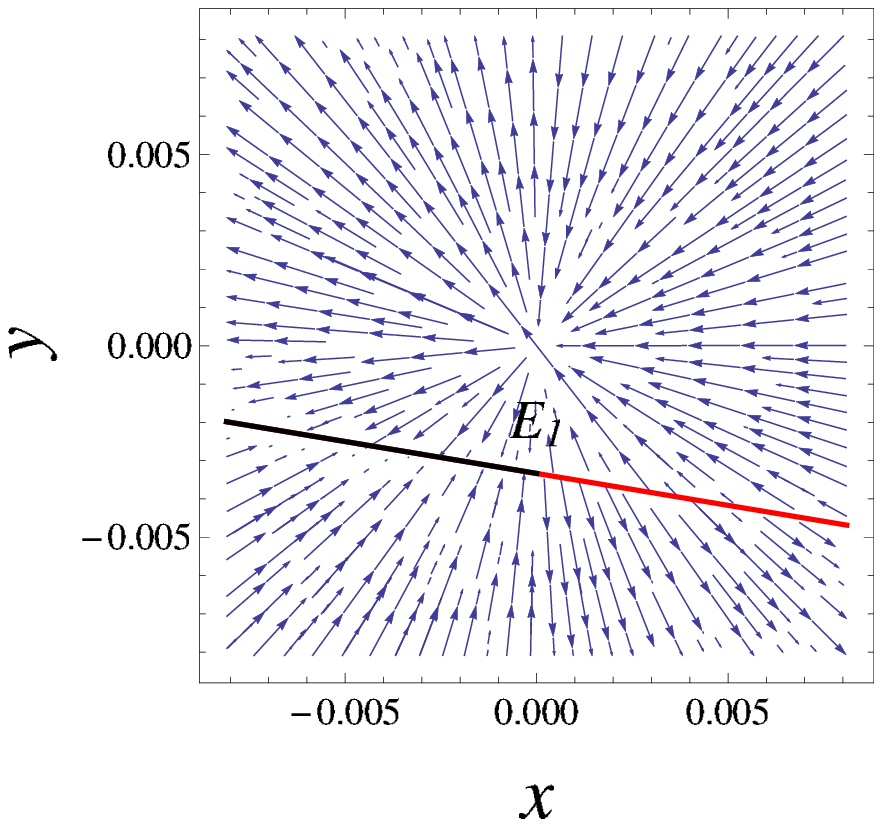}\label{fig:stable_E1}}
\caption{Figure shows the vector field of autonomous system (\ref{Int1de1}-\ref{Int1de2}) for the interaction model 1 with the parameter values $\alpha_{m}=-0.01$ and $ \Gamma_{0}=0.5 $. In panel (a), a colored line represents a line $A_1$ with black colored line being a stable portion of set $A_1$ and in panel (b), a colored line represents a line $E_1$ with black colored line being a stable portion of set $E_1$.}
\label{fig:stable_A1_E1}
\end{figure}

The system (\ref{Int1de1})-(\ref{Int1de2}) admits two sets of critical points $A_1$ and $E_1$. As mentioned earlier point $B_1$ lies on the set $E_1$, whereas points $C_1$, $D_1$ lie on the set $A_1$. In what follows, we therefore analyze the stability of sets $A_1$ and $E_1$ only. \\
$\bullet $  The solution associated with the set of critical points $A_1 (1,y_c)$ (where $y_c$ takes any real value) always exist.
They are completely DE dominated solutions ($\Omega_{d}=1$), where DE corresponds
to an exotic type fluid with equation of state $\omega_{d}=y_c$. For
this case, DE can describe the quintessence, cosmological constant, or
phantom field, or any other perfect fluid according as the choice
of $y_c$. So, the line of critical points may have different features in their
cosmic evolution.  Points on this set corresponds to an accelerating universe ({\it i.e.}, $\omega_{\rm eff}<-\frac{1}{3}$)
 for $y_c<-\frac{1}{3}$ (see Table \ref{modelQ1}), and there exists an
expanding universe if the evolution of Hubble function satisfies $\omega_{\rm eff}<-1$,
( see Eq. (\ref{Hubble-fn})) ({\it i.e.}, Hubble parameter increases gradually)
for $y_c<-1$, {\it i.e.}, in phantom region. This set is normally hyperbolic and hence corresponds to a late time attractor for  $y_c< \frac{\alpha_{m}-\Gamma_{0}}{3}$ (see Table \ref{evalues2D}). This is also being confirmed numerically in Fig.\ref{fig:stable_A1}. The one dimensional center subspace spanned by eigenvector \[\left(\begin{array}{c}
0\\
1\end{array} \right)\] which corresponds to a vanishing eigenvalue
identifies the direction of the set $A_1$. Whereas, the one
dimensional stable subspace near this set is spanned by
eigenvector \[\left(\begin{array}{c}
1/y_c\\
1\end{array} \right)\] which corresponds to a non-vanishing eigenvalue with $y_c< \frac{\alpha_{m}-\Gamma_{0}}{3}$.
 Since there is no unstable subspace near $A_1$ when $y_c< \frac{\alpha_{m}-\Gamma_{0}}{3}$, so trajectories will approach
 towards some points on set $A_1$. While the set of critical points $A_1$ represents a stable attractor in quintessence
 region for $y_{c}<{\rm min} \left\{ \frac{\alpha_{m}-\Gamma_{0}}{3}, -\frac{1}{3} \right\}$, they are stable solutions with cosmological constant behavior
for $\Gamma_{0}<3+\alpha_{m}$. On the other hand, stable solutions
are obtained in the phantom region for $y_c< {\rm min}
\{-1,\frac{\alpha_{m}-\Gamma_{0}}{3} \}$. Hence, the set of
critical points represents the solutions of accelerated stable
attractor in some parameter region where DE behaves as
quintessence, cosmological constant, or phantom field. This is one
of the important results in this context of interacting DE since
in this scenario DE can mimic three distinct phases of the cosmic
evolution. It should be noted that in Fig. \ref{fig:stable_A1_E1}
the origin ($0,0$) of the phase space acts as a critical point. As
the dynamical system is singular at this point so its stability can not
be determined directly, only numerical investigation can infer its behavior.
So, we can say that ($0,0$) acts as a (non-linear) critical point of the
system.


$\bullet $  The set $E_1$ exists for all model parameters. It represents a scaling solution where DM and DE
scale as $\Omega_{m}/\Omega_{d}=(1-x_c)/ x_c$. The DE describes any perfect fluid with equation of state parameter
$\omega_{d}=\frac{\alpha_{m}-\Gamma_{0} x_c}{3x_c}$. This set is normally hyperbolic and hence, it is stable when
 $0<x_c<1$, $\alpha_m>0$; $x_c<0$, $\alpha_m<0$; $x_c>1$, $\alpha_m<0$. This is confirmed numerically in Fig. \ref{fig:stable_E1}.
 The one dimensional stable subspace near this set is spanned by eigenvector \[\left(\begin{array}{c}
\frac{3x}{\alpha_m-\Gamma_0 x}\\
1\end{array} \right).\] with $x_c$ and $\alpha_m$ satisfying the above stability condition. The eigenvector  \[\left(\begin{array}{c}
-\frac{3}{\Gamma_0}\\
1\end{array} \right)\] which corresponds to vanishing eigenvalue determines the direction of the set.
 This set describes an accelerated quintessence behavior for $1<\Gamma_{0}-\alpha_{m}<3$ ($-1<\omega_{\rm eff}<-\frac{1}{3}$),
  while it represents a cosmological constant behavior for $\Gamma_{0}=3+\alpha_{m}$ ($\omega_{\rm eff}=-1$), and phantom behavior for
$\Gamma_{0}>3+\alpha_{m}$ ($\omega_{\rm eff}<-1$, see table
\ref{modelQ1}). This set is interesting from the cosmological
point of view as it describes late time attractor in quintessence,
cosmological constant or in phantom region for some choice of
$\alpha_m$ and $x_c$. Interestingly, from Fig. \ref{fig:stable_A1_E1} the origin $(0,0)$ behaves as a (non-linear) critical point of the system. However, at this point the system is singular and hence its stability cannot be determined analytically, but numerically the system behaves as if the origin is not stable.

\begin{figure}
\centering
\includegraphics[width=6cm,height=4cm]{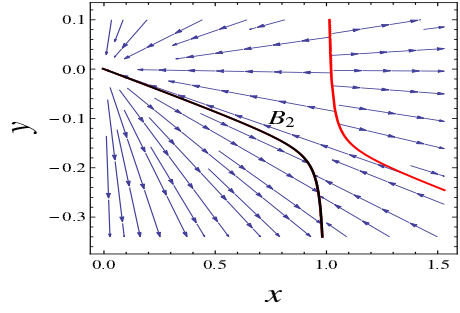}
\caption{Figure shows the vector field of autonomous system (\ref{Int2de1}-\ref{Int2de2}) for the interaction model 2 with the parameter values $\alpha_{d}=-0.01$ and $ \Gamma_{0}=0.5 $. A colored curve represents curve $B_2$ with black colored curve being a stable portion of set $B_2$.}
\label{fig:stable_B2}
\end{figure}


\subsection{Interaction 2}

$\bullet $  The autonomous system (\ref{Int2de1})-(\ref{Int2de2})
has only one set critical points $B_2$. As mentioned earlier, it
is to be noted that point $A_2$ lies on the set $B_2$. So, we
shall analyze the stability of set $B_2$ only. Set $B_2$
corresponds to a scaling solution and it always exist except $x_c
= 1$. For this solution, DM and DE scale in a constant fraction
as: $\Omega_{m}/\Omega_{d}=(1-x_c)/x_c$, where DE behaves as a
perfect fluid having barotropic equation of state
$\omega_{d}=\frac{\Gamma_{0}x_c-\Gamma_{0}+\alpha_{d}}{3(1-x_c)}$
(see table \ref{modelQ2}). This set corresponds to an accelerated
universe if $\frac{\alpha_d\,x_c}{1-x_c}<\Gamma_0-1$.
 This set is normally hyperbolic and hence it is stable if  $\alpha_{d}<0$, $0<x_c<1$; $x_c<0$, $\alpha_d>0$; $x_c>1$, $\alpha_d>0$. This is confirmed numerically from Fig. \ref{fig:stable_B2}.  The eigenvector  \[\left(\begin{array}{c}
-\frac{3(x_c-1)^2}{\Gamma_0(x-1)^2-\alpha_d}\\
1\end{array} \right)\] corresponds to vanishing eigenvalue determines the direction of the tangent at each point of the set. Whereas the one dimensional stable subspace near this set is spanned by eigenvector \[\left(\begin{array}{c}
\frac{3(1-x)}{\Gamma_0 (x-1)+\alpha_d}\\
1\end{array} \right)\] with $\alpha_d$ and $x_c$ satisfying the
above stability condition. So, depending on the choice of
$\alpha_d$ and $x_c$ this set can explain the late time behavior
of our universe.


\begin{figure}
\centering
\includegraphics[width=6cm,height=4cm]{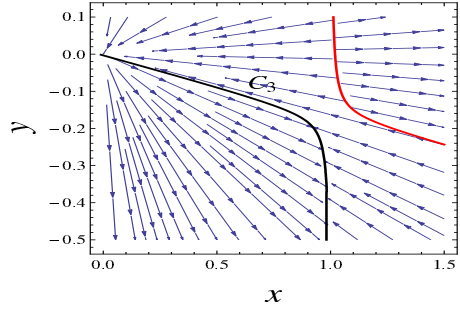}
\caption{Figure shows the vector field of autonomous system (\ref{Int3de1}-\ref{Int3de2}) for the interaction model 3 with the parameter values $\alpha=-0.01$ and $ \Gamma_{0}=0.5 $.  A colored curve represents curve $C_3$ with black colored curve being a stable portion of set $C_3$.}\label{fig:stable_C3}
\end{figure}

\subsection{Interaction 3}
$\bullet $ The system (\ref{Int3de1})-(\ref{Int3de2}) admits only
one critical set of points $C_3$. As mentioned earlier, we see
that points $A_3$, $B_3$ are points on the set $C_3$. So, we shall
analyse the stability of set $C_3$ only. One interesting point for
this solution $C_3$ is that it is a combination of DM and DE
having the ratio:
$\frac{\Omega_{m}}{\Omega_{d}}=\frac{1-x_c}{x_c}$, and will exist
for all model parameters except $x_c = 1$. The accelerating
universe is predicted by the set when
$\frac{\alpha}{1-x_c}<\Gamma_0-1$. This set is normally hyperbolic
and it is stable if $0<x_c<\frac{1}{2}$, $\alpha>0$;
$\frac{1}{2}<x_c<1$, $\alpha<0$; $x_c<0$, $\alpha<0$; $x_c>1$,
$\alpha>0$. Hence, this set provides some interesting features for
positive coupling as well as negative coupling of interaction. Its
stability is being confirmed numerically in Fig.
\ref{fig:stable_C3}. The one dimensional stable subspace near this
set is spanned by eigenvector \[\left(\begin{array}{c}
\frac{3 x_c(1-x_c)}{\Gamma_0 x_c(x_c-1)+\alpha}\\
1\end{array} \right)\] which corresponds to a nonvanishing eigenvalue with $\alpha$, $x_c$ satisfying the above stability condition. Whereas, the direction of the tangent at each point on the set is along the eigenvector
 \[\left(\begin{array}{c}
\frac{3(x_c-1)^2}{\alpha-\Gamma_0 (x_c-1)^2}\\
1\end{array} \right)\] corresponds to a vanishing eigenvalue. So,  depending on the choice of parameters and fine tuning of initial conditions,
trajectories near this set approach towards points of this set. Hence, some critical points on this set corresponds to a late time accelerated universe.\\


\begin{figure}
\centering
\includegraphics[width=6cm,height=4cm]{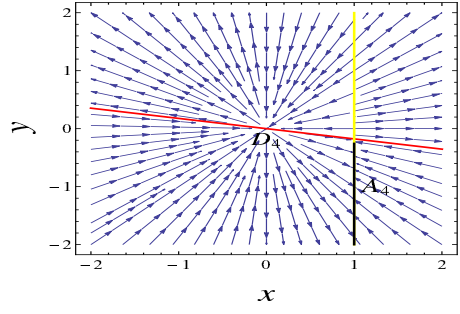}
\caption{Figure shows the vector field of autonomous system (\ref{Int4de1}-\ref{Int4de2}) for the interaction model 4 with the parameter values $\beta=-0.01$ and $ \Gamma_{0}=0.5 $. Colored lines represent line $A_4$ and $D_4$ with black colored curve being a stable portion of set $A_4$.}
\label{fig:stable_A4_D4}
\end{figure}

\subsection{Interaction 4}
There are two sets of critical points arising from the interaction
model 4 ($A_4$ and $D_4$). As mentioned earlier, points $B_4$,
$C_4$ are points on the set $A_4$. So, in what follows  we shall
analyse the stability of set $A_4$ and $D_4$ only

$\bullet $  The set of critical points $A_4$ exists for all model
parameters involved. It represents a DE dominated solution
($\Omega_{d}=1$). This DE dominated solution describe the late
time acceleration of universe when DE behaves as quintessence,
cosmological constant or phantom or any other exotic fluid for
$y_{c}<-\frac{1}{3}$. This set is again normally hyperbolic and it
is stable when $y_c< \frac{\alpha_{m}-\Gamma_{0}}{3}$. The
stability of $A_4$ is confirmed numerically in Fig.
\ref{fig:stable_A4_D4}. The one dimensional stable subspace near
this set is spanned by eigenvector \[\left(\begin{array}{c}
1/y_c\\
1\end{array} \right)\] where  $y_c< \frac{\alpha_{m}-\Gamma_{0}}{3}$. The eigenvector \[\left(\begin{array}{c}
0\\
1\end{array} \right)\] corresponds to vanishing eigenvalue
determines the direction of a set. Since, there is no unstable
subspace near this set for $y_c< \frac{\alpha_{m}-\Gamma_{0}}{3}$.
This means that depending on choice of $\alpha_m$, $\Gamma_0$ and
$y_c$ trajectories approach towards some points on this set.

$\bullet $ The solution represented by the point $D_4$ is combination of
both DE and DM with the constant ratio as:
$\frac{\Omega_{m}}{\Omega_{d}}=\frac{1-x_c}{x_c}$, where DE describes any perfect fluid
with equation of state $\omega_{d}=\beta-\frac{\Gamma_{0}}{3}$. The set exists for all model parameters. Depending on some parameter restrictions, an acceleration will occur
for the set, but since  both of eigenvalues vanish, we do not obtain any information regarding the stability of $D_4$ (see Fig. \ref{fig:stable_A4_D4}). It behaves as a neutral line.


\begin{figure}
\centering
\subfigure[]{%
\includegraphics[width=6cm,height=4cm]{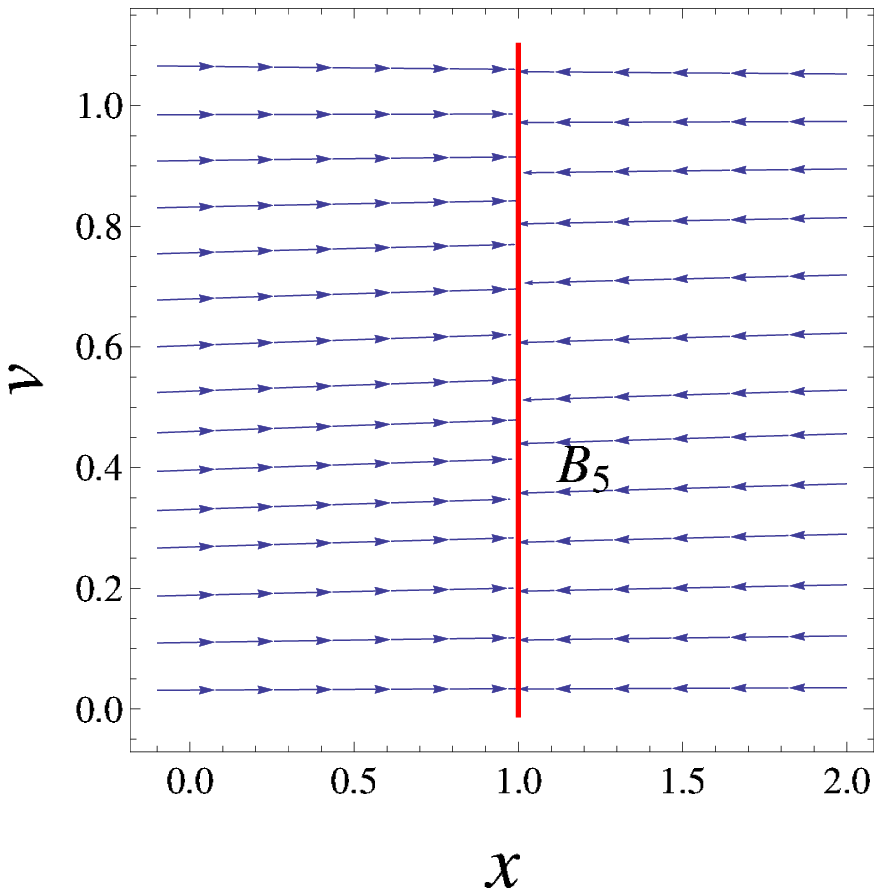}\label{fig:fig_xv_B5}}
\qquad
\subfigure[]{%
\includegraphics[width=6cm,height=4cm]{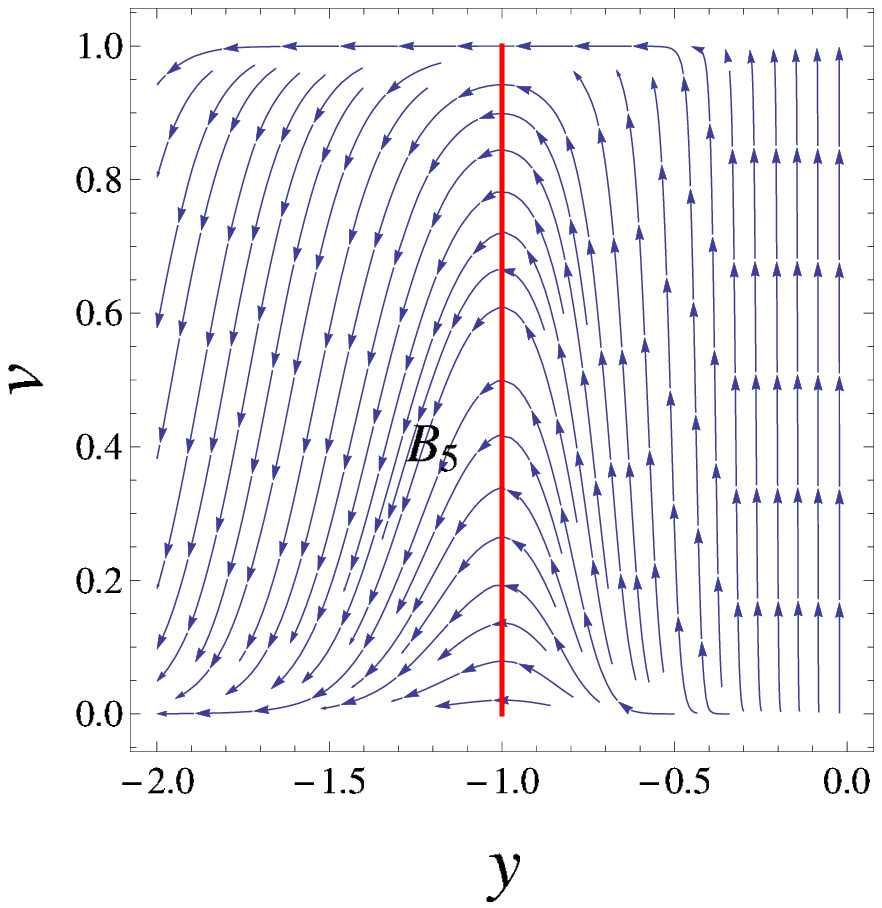}\label{fig:fig_yv_B5}}
\caption{Figure shows the vector field projection on (a) $xv$ phase plane and (b) $yv$ phase plane of autonomous system (\ref{Int5de1}-\ref{Int5de3}) for the interaction model 5 with the parameter values $\gamma=0.5$ and $ \Gamma_{0}=0.001$.}
\label{fig:B5}
\end{figure}

\begin{figure}
\centering
\includegraphics[width=6cm,height=4cm]{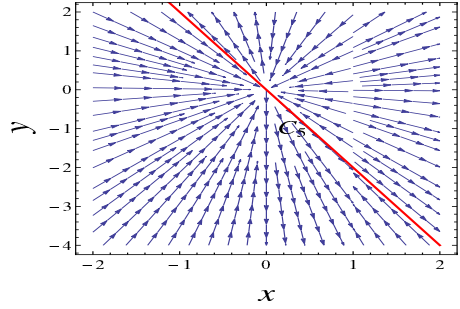}
\caption{Figure shows the vector field projection on  $xy$ phase plane of autonomous system (\ref{Int5de1}-\ref{Int5de3}) for the interaction model 5 with the parameter values $\gamma=1$ and $ \Gamma_{0}=6$. It may be noted that we take $\Gamma_{0}$ to be very large for this particular plot simply to check its unstability, since for $\Gamma_0<3$, eigenvalue $\lambda_3>0$ and it will be surely unstable.}
\label{fig:C5}
\end{figure}

\subsection{Interaction 5}
$\bullet $ From the local interaction model-5, we get four sets of
critical points presented in table \ref{modelQ5}. The set $A_5$
(similar with $A_1$ and $A_4$ ) is completely DE dominated
solution. It corresponds to phantom universe for $y_c <-1$, and it
corresponds to a quintessence dominated phase for
$-1<y_c<-\frac{1}{3}$. The DE associated with this set can mimic
any kind of fluid for different choices of $y_c$. It is a normally
hyperbolic set and hence it is stable if
$(a)~\Gamma_{0}\leq3~\mbox{and}~y_c<-1$, or
$(b)~\Gamma_{0}>3~\mbox{and}~y_c<-\frac{\Gamma_{0}}{3}$. This
means, that the set will be stable only in the phantom regime.
Furthermore, in this region the set becomes physically relevant
describing the late time accelerated expansion of the universe.
The two dimensional stable subspace is spanned by eigenvectors
\[\left(\begin{array}{c}
1/y_c\\
1\\
0\end{array} \right) \text{and} \left(\begin{array}{c}
0\\
0\\
1\end{array} \right)\] corresponds to two nonvanishing eigenvalues where $y_c$ satisfies the above stability condition. The one dimensional center subspace spanned by eigenvector \[ \left(\begin{array}{c}
0\\
1\\
0\end{array} \right)\] corresponds to a vanishing eigenvalue and determines the direction of set $A_5$. Hence, this set can describe the late time behavior of our universe.\\

$\bullet $ The set of points $B_5$ exists for all values of model parameters. This solution is completely
DE dominated, where DE behaves as cosmological constant. There exists always
an accelerating universe ($\omega_{\rm eff}=-1$, see table \ref{modelQ5}). It is a non-isolated set of critical points
 where all points are nonhyperbolic but it is not normally hyperbolic set since it contains two vanishing eigenvalues.
  Numerical projection plot of the system (\ref{Int5de1})-(\ref{Int5de3}) shows that this set cannot be stable. It can be seen that in $(x,v)$ phase space,
  trajectories are attracted toward the set $B_5$ (see Fig. \ref{fig:fig_xv_B5}), however trajectories in $(y,v)$ phase space are not
  attracted towards the set $B_5$ (see Fig. \ref{fig:fig_yv_B5}). We have checked that this actually happens for different choices of
   model parameters. This implies that points of this set are saddle in nature.


$\bullet $  The set of critical points $C_5$ exists for all values of model parameters. The
set corresponds to a solution in combination with both DE and DM in the phase space with DE behaves as any perfect fluid model  having
equation of state parameter  $\omega_{d}=-\Gamma_{0}/3$. Hence, it is clear that
the DE may have different features during its evolution such as quintessence is
described by DE in the parameter region $1<\Gamma_{0}<3$, while DE describes cosmological
constant for $\Gamma_{0}=3$, and phantom regime is realized for $\Gamma_{0}>3$.
Now the expansion of the universe is accelerated for  $\Gamma_{0}>1$ ($\omega_{\rm eff}<-\frac{1}{3}$).
 This set is again non-hyperbolic but not normally hyperbolic. Numerically, by plotting the projection of trajectories  on $(x,y)$ plane (see Fig. \ref{fig:C5}), we observe that points on this set are saddle in nature.


$\bullet $ The set of points $D_5$  is the combination of both DE
and DM. This set behaves as a cosmological constant ({\it i.e.},
$\omega_{\rm eff}= -1$, see table \ref{modelQ5}), and hence there
is always an accelerating universe near this set. Also, the set of
critical points under consideration is a normally hyperbolic set.
Hence, it is stable spiral if $\Gamma_0<3$, $x_c>1$ or
$\Gamma_0<3$, $x_c<-1$, it is stable node if $-1<x_c<0$,
$\Gamma_0<3$ or $0<x_c<1$, $\Gamma_0>3$.


\section{Cosmological Implications}
\label{cosmological-implications}

\begin{figure}
\centering
\subfigure[]{%
\includegraphics[width=6cm,height=4cm]{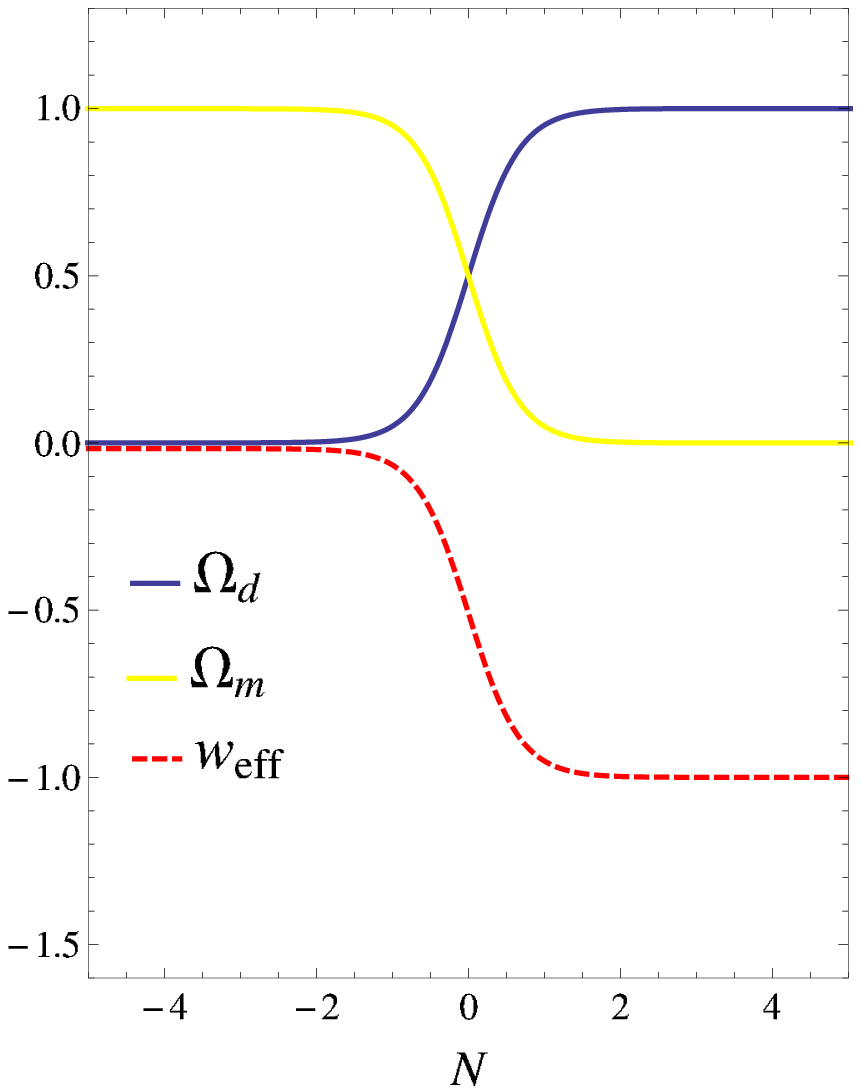}\label{fig:parameters1_L}}
\qquad
\subfigure[]{%
\includegraphics[width=6cm,height=4cm]{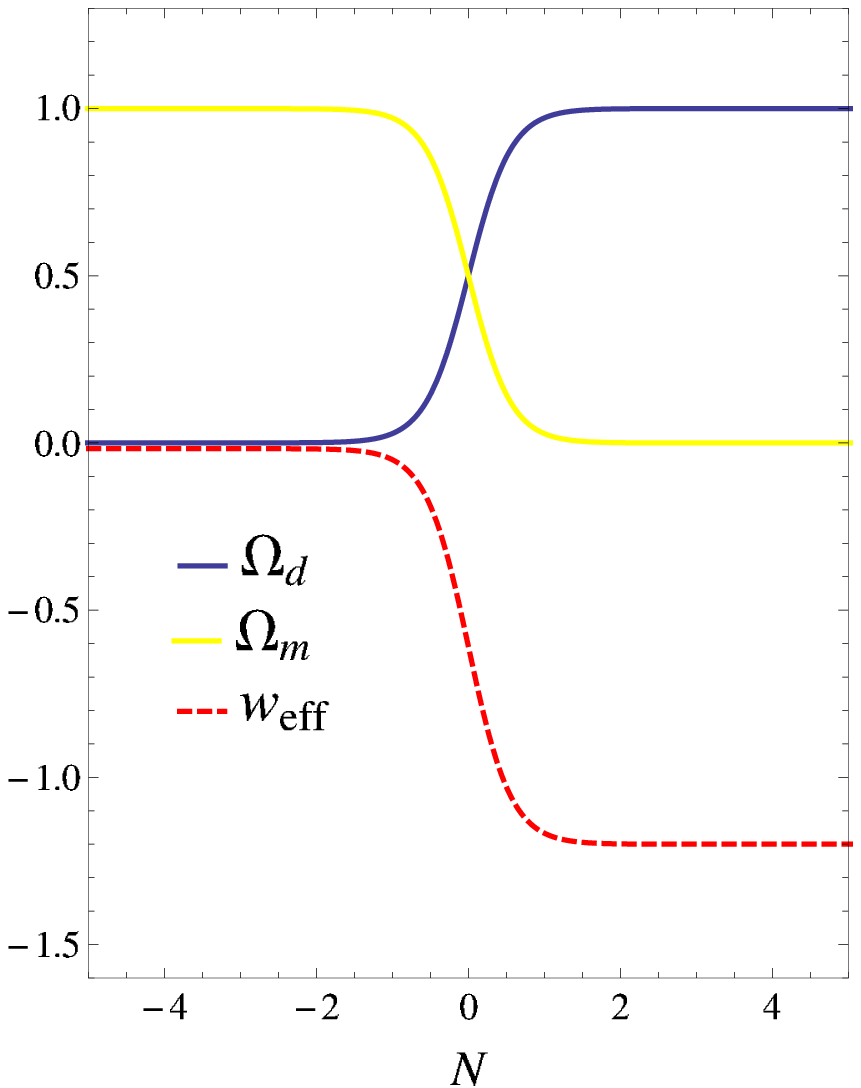}\label{fig:parameters1_P}}
\caption{Figure shows the evolution of DE energy density parameter $\Omega_d$, DM energy density parameter $\Omega
_m$ and effective equation of state $\omega_{\rm eff}$ of the system (\ref{Int1de1}-\ref{Int1de2}) for the interaction model 1 with the parameter values $\alpha_{m}=-0.001,~~and~~ \Gamma_{0}=0.05$ with different choices of initial conditions. Panel (a) shows the cosmological constant as late time attractor and panel (b) shows the phantom regime as late time attractor.}
\label{fig:parameters}
\end{figure}
In this section, we shall describe the main cosmological features extracted
from the interacting dark energy models in presence of gravitational particle production.
In the following subsections, we shall describe the physics of the critical points for each
interacting model in this framework along with their viability to describe different cosmic phases. An interesting feature is that in all the interaction models, the evolution of $\Omega_m$, $\Omega_d$ and $\omega_{\rm eff}$ (see Fig. \ref{fig:parameters}) are similar and so we have not plotted for each interaction model.

\subsection{Interaction Model 1:}

In this model, we obtained two set of critical points $A_1$ and
$E_1$. Set $A_1$ represents a de-Sitter universe for $y_c=-1$
(point $D_1$), it represents a stiff matter dominated universe for
$y_c=1$ and for $y_c=0$ ({\it i.e.}, point $B_1$) we actually get
a DE dominated universe ($\Omega_d=1$) but it appears as if it was
matter dominated solution ($\omega_{\rm eff}=0$). It is also noted
that critical point $B_1$ is special case of set $E_1$ and it
corresponds to a matter dominated universe for $\alpha_m= 0$ when
no interaction between DE and DM is considered. Moreover, set
$E_1$ also represents a matter dominated universe for $x_c=0$
({\it i.e.}, point $(0,\frac{\alpha_m}{3})$). From the analysis
performed in Sec. \ref{phase-space}, we see that depending on the
choice of model parameters and fine tuning of initial conditions,
the Universe evolves from a matter dominated phase (set $E_1$) to
a DE dominated phase (set $A_1$), either to a quintessence regime
for $-1<y_c<-\frac{1}{3}$, or cosmological constant for $y_c=-1$,
or phantom regime for $y_c<-1$ (see for e.g., Fig.
\ref{fig:parameters}). Hence, we observe that the background
dynamics of this model can possibly mimic the $\Lambda$CDM model
(see Fig. \ref{fig:parameters1_L}). Moreover, there is a
possibility of crossing the phantom barrier ($\omega_{\rm
eff}=-1$, see Fig. \ref{fig:parameters1_P}) which is slightly
favored by observations and cannot be achieved in case of
non-interacting DE models. Hence, this model can well describe the
late time transition from DM to DE dominated phase of the
universe.
\subsection{Interaction Model 2:}

In this model, there is only one set of critical points $B_2$. This set represents a DM dominated universe when $x_c=0$, {\it i.e.},
when we consider the origin $(0,0)$ as a critical point. Unfortunately, we do not obtain any information regarding the stability of point
$(0,0)$ as both eigenvalues vanishes for this particular point. However, from Fig. \ref{fig:stable_B2}, it looks like point $(0,0)$ is not
 stable. Critical point $A_2$ is a special case of set $B_2$ and it represents a DE dominated but decelerated universe ($\Omega_d=1$, $\omega_{\rm eff}=0$)
  for $\alpha_d = 0$ when there is no coupling between DE and DM.
   Set $B_2$ can represent a late time accelerated scaling solution for $\alpha_{d}<0$, and $0<x_c<1$. However,
    viable trajectories are attracted towards $B_2$ near the limit $x_c=1$, {\it i.e.}, DE dominated universe
    (see Fig. \ref{fig:stable_B2}). So, for this model the Universe evolves from a matter dominated solution (set $B_2$ for $x_c=0$)
     towards a DE dominated solution (set $B_2$ for limit $x_c \rightarrow 1$) (a similar scenario is obtained for this model as in Fig. \ref{fig:parameters}).

\subsection{Interaction Model 3:}

The background cosmological behavior of this model is similar with
interaction model 2. In this model, there is only one set of
critical points $C_3$. This set represents a DM dominated universe
when $x_c = 0$ for which no information is obtained regarding its
stability as both eigenvalues vanishes for this particular
  case. However, numerically it can be seen from Fig. \ref{fig:stable_C3} that this set is not stable. Critical points $A_3$ and $B_3$
  are
   special cases of set $C_3$ and correspond to  a scaling solutions. Point $A_3$ corresponds
    to DE dominated but decelerated universe ($\Omega_d=1$, $\omega_{\rm eff}=0$) for $\alpha= 0$. So, when no interaction is considered this point corresponds to a DE dominated universe
    but the universe expands as if it was matter dominated. Point $B_3$ corresponds to a DM
      dominated universe for $\alpha=0$. Set $C_3$ can represent a late time accelerated scaling solution for some choices of $\alpha$ and some $x_c$.
      It corresponds to a DM dominated universe for $x_c=0$. Moreover, viable trajectories are attracted towards $C_3$ near the limit $x_c=1$ (see Fig. \ref{fig:stable_C3}).
       So, for this model the Universe evolves from a matter dominated solution (set $C_3$ for $x_c=0$) towards a DE dominated solution (set $C_3$ for $x_c \rightarrow 1$).

\subsection{Interaction Model 4:}

In this model, we obtained two sets of critical points $A_4$ and
$D_4$. Set $A_4$ represents a stiff matter dominated universe for
$y_c=1$, it also represents a de-Sitter universe for $y_c=-1$
(\textit{i.e.,} point $C_4$). For $y_c=0$ ({\it i.e.}, point
$B_4$), this set corresponds to a DE dominated universe
($\Omega_d=1$), but the universe appears as if it was matter
dominated ($\omega_{\rm eff}=0$). Set of critical points $D_4$
behaves as a neutral line, its stability cannot be determined as
all eigenvalues vanishes. However, for $x_c=0$, {\it i.e.}, point
$(0,0)$, it corresponds to a matter dominated universe. Even
though stability cannot be determined analytically,
    numerically it shows that the origin is not stable (see Fig. \ref{fig:stable_A4_D4}). Hence, we see that depending
     on the choice of model parameters and fine tuning of initial conditions, the Universe evolves from a matter
      dominated phase (set $D_4$) to a DE dominated phase (set $A_4$) either to a quintessence regime for $-1<y_c<-\frac{1}{3}$,
       or cosmological constant for $y_c=-1$, or phantom regime for $y_c<-1$.

\subsection{Interaction Model 5:}

In this model, we obtained four sets of critical points ($A_5$,
$B_5$, $C_5$ and $D_5$). Set $A_5$ corresponds to a late time
attractor where DE dominates only in phantom regime. Critical
 points on sets $B_5$ and $C_5$ behaves as saddle and interestingly set $C_5$ corresponds
 to a matter dominated universe for $x_c=0$. Set $D_5$ also corresponds to late time accelerated
  solution for some choices of model parameters. Hence, depending on initial conditions
  and choices of model parameters we see that the universe can evolves from matter dominated phase (set $B_5$ or $C_5$)
   to a DE dominated phase (set $D_5$). Physically, it means that there is a transition from DM to DE domination in late universe.

\section{Short Discussion}
\label{short-discussion}

In the present work, we have performed a dynamical system analysis
for the scenarios of interacting dark matter and dark energy,
where additionally the gravitational particle production is also
allowed. The particle creation mechanism describes many
interesting results such as the possibility of phantom universe
without invoking any phantom field, formation of emergent
universe, complete cosmic scenario etc. Here, we have considered
the dark matter fluid as dust and the dark energy as a perfect
fluid with equation of state $\omega_d$. Moreover, the created
particles by the gravitational field have been considered to be
dark matter particles (equivalently, dust particles) in agreement
with the local gravity constraints and the production rate is
taken to be varying linearly with the Hubble function ({\it i.e.},
$\Gamma \propto H$). We have considered five interacting models
which  correspond to five distinct forms of interaction $Q$. The
objective for choosing such a complex system is to examine whether
there is any model (interacting) which could explain the overall
evolution of the universe. In particular, a complete description
of evolution at late times can be obtained in quintessence,
$\Lambda$CDM, or phantom era connected through a DM dominated era.
Critical points, their existence, and their corresponding
cosmological parameters are shown in tables \ref{modelQ1} $-$
\ref{modelQ5} for the respective models. Additionally, we have
presented the eigenvalues for different interaction models in
tables \ref{evalues2D} and \ref{evalues3D}. A detailed stability
analysis has been executed successively in Sec. \ref{phase-space}.
It is also noted that all sets of points except $D_4$, $B_5$,
$C_5$ are normally hyperbolic, where stability is confirmed by the
signature of the remaining non-vanishing eigenvalues.

We found that the sets of critical points ($A_1$, $A_4$, $A_5$)
corresponds to DE dominated universe where DE could mimic
quintessence era, cosmological constant, phantom phase, sometimes
dust or even any other exotic fluid. However, it was found that
some of the critical points in the above set of critical points
representing the above cosmic phases ({\it i.e.}, quintessence,
cosmological constant, or phantom phase) could describe the
late-time expansion of the universe but they can not alleviate the
coincidence problem. On the other hand, some sets of critical
points ($E_1$, $B_2$, $C_3$, $D_5$) can possibly represent scaling
solutions for $0<x_c<1$ with an accelerated expansion of the
universe. However, we observe that trajectories are attracted
towards a portion of sets where $x_c \approx 1$ and hence critical
points of these sets cannot alleviate the coincidence problem
(since $\Omega_d \approx 1$). Moreover, critical points on sets
$E_1$, $B_2$, $C_3$, $D_5$ with $x_c=0$ represents a DM dominated
universe. Stability analysis (in Sec. \ref{phase-space}) shows
that for some choices of model parameters and fine tuning of
initial conditions, one can connect these DM dominated solutions
to a DE dominated solutions ($A_1$, $A_4$, $A_5$ or $B_2$, $C_3$
for limit $x_c \rightarrow 1$) which can possibly mimic
quintessence, cosmological constant, or phantom phase. It may be
noted that phenomenological interesting solutions depends on the
strong fine tuning of the initial conditions since all the
critical points lie on different non-isolated sets and only few
points describe the correct observed cosmological dynamics.

Thus, in summary, one may conclude that the present interacting DE model in the framework
of particle creation mechanism may describe different evolutionary phases of the universe.
 These interacting models can possibly allow the crossing of phantom divide line (see Fig. \ref{fig:parameters1_P}
  which shows the clear cosmic evolution of the physical quantities $\omega_{\rm eff}$, $\Omega_{d}$, $\Omega_{m}$)
   which is not possible in case of uncoupled standard cosmology. The present particle creation mechanism describes the
   true non-equilibrium thermodynamics of the universe compared to others standard DE models. As a result, present model
   shows stable critical points representing various cosmological scenarios. Moreover, the background dynamics of these interacting
    models can possibly mimic the $\Lambda$CDM but only for $y_c=-1$, so there might be some differences at the level of perturbations.
    However, cosmological perturbation analysis lies beyond the scope of our present study. This can be left for future works.

\section*{Acknowledgments}

The authors are grateful to the referee for some critical and useful comments to improve the work. SC thanks IUCAA for providing research facilities and also the UGC-DRS programme, at the department of Mathematics, Jadavpur University. SKB would like to thank S. Pan for helpful discussions on this work, and also acknowledges IUCAA for providing research facilities in the library where a part of the work was done during a visit.

\end{document}